\documentclass[%
 reprint,
 amsmath,amssymb,
 aps,
 pra,
]{revtex4-2}

\usepackage{graphicx}
\usepackage{dcolumn}
\usepackage{bm}
\usepackage{braket}

\usepackage{amsmath,amsthm}
\usepackage{dsfont}
\usepackage{mathtools}
\usepackage{framed}
\usepackage{siunitx}
\usepackage{comment}
\usepackage{makecell}
\usepackage{tabularx}
\usepackage{amssymb}
\usepackage{hyperref}

\newcounter{protcounter}
\newenvironment{prot}
{
\refstepcounter{protcounter}
\begin{framed}
\noindent \textbf{Protocol~\theprotcounter:}\\}
{\end{framed}}

\usepackage{color}
\usepackage{xcolor}

\begin{document}

\preprint{APS/123-QED}

\title{Gigabit-rate Quantum Key Distribution on Integrated Photonic Chips}
\author{Si Qi Ng$^1$}
\author{Florian Kanitschar$^{2,3}$}
\author{Gong Zhang$^4$}
\author{Chao Wang$^4$}
\email{wang.chao@nus.edu.sg}
\affiliation{$^1$Centre for Quantum Technologies, National University of Singapore, Singapore}
\affiliation{$^2$Vienna Center for Quantum Science and Technology (VCQ), Atominstitut, Technische Universität Wien, Stadionallee 2, 1020 Vienna, Austria}
\affiliation{$^3$AIT  Austrian  Institute  of  Technology,  Center  for  Digital  Safety\&Security,  Giefinggasse  4,  1210  Vienna, Austria}
\affiliation{$^4$Department of Electrical \& Computer Engineering, National University of Singapore, Singapore }%

\date{\today}

\begin{abstract}

Quantum key distribution (QKD) provides information-theoretic security guaranteed by the laws of quantum mechanics, making it resistant to future computational threats, including quantum computers.
While QKD technology shows great promise, its widespread adoption depends heavily on its usability and viability, with key rate performance and cost-effectiveness serving as critical evaluation metrics. 
In this work, we report an integrated silicon photonics-based QKD system that achieves a secret key rate of 1.213 $\SI{}{\giga\bit\per\second}$ over a metropolitan distance of 10 km with polarization multiplexing. 
Our contributions are twofold. First, in the quantum optical layer, we developed an on-chip quantum transmitter and an efficient quantum receiver that operate at 40 Gbaud/s at room temperature.  
Second, we designed a discrete-modulated continuous variable (DM CV) QKD implementation with efficient information reconciliation based on polar codes, enabling potentially high-throughput real-time data processing. 
Our results demonstrate a practical QKD solution that combines high performance with cost efficiency. We anticipate this research will pave the way for large-scale quantum secure networks. 

\end{abstract}

\maketitle
\section{Introduction}\label{sec:introduction}

Information security serves as a fundamental pillar of our digital infrastructure, providing critical services, including data confidentiality, integrity, and authenticity. These cryptographic functions play a crucial role in activities ranging from national security and defense to everyday Internet browsing and financial transactions. 
As its core, secure key distribution presents an essential cryptographic primitive that enables remote parties (typically referred to as Alice and Bob) to establish shared random keys while maintaining security against potential adversaries (Eve). 
However, the emergence of quantum computers, along with rapid advances in classical computing technology, poses a significant challenge to the security of current cryptographic protocols that rely on the computational intractability of specific mathematical problems. 

Quantum key distribution (QKD) is a protocol whose security relies on the laws of quantum mechanics rather than computational complexity. This foundation gives QKD distinct advantages over conventional cryptographic protocols, providing long-term security that is resistant to threats from supercomputers and quantum computers, regardless of future technological advancements~\cite{2002_gisin_QuantumCryptography,2009_scarani_SecurityPracticalQuantum,2020_pirandola_AdvancesQuantumCryptography,2020_xu_SecureQuantumKey, usenko2025continuousvariablequantumcommunication}. 

This is important not only when attacks are imminent but also before powerful decryption methods emerge. Indeed, an adversary can store encrypted messages today and decrypt them using future technology, a strategy known as the ``acquire-now-decrypt-later'' attack. In this context, QKD offers an attractive solution, particularly for sensitive information that must remain secure for decades or even a lifetime.

The development of QKD has seen tremendous progress in both theory and experiment in the last four decades~\cite{2020_xu_SecureQuantumKey,2020_pirandola_AdvancesQuantumCryptography,2022_portmann_SecurityQuantumCryptography}. 
For practical applications, improving the technology's usability and viability represents a critical priority. This necessitates the development of high-performance, cost-effective QKD solutions to enable large-scale deployment. 

The secret key rate (SKR), which measures the number of secret key bits that Alice and Bob can generate per unit time, is an essential performance metric for QKD systems. It is particularly critical for encryption schemes that aim for the highest level of security, where encryption keys must be as long as the message they protect~\cite{1949_shannon_CommunicationTheorySecrecy}. For example, when considering secure storage of genomic data for thousands of individuals, it has been suggested that QKD systems operating at gigabit per second rates are needed for One-Time-Pad (OTP) encryption for lifelong protection~\cite{2017_sasaki_QuantumNetworksWhere}.

Achieving high-rate QKD presents significant challenges, as it requires both the quantum optics layer and classical data post-processing layer to operate efficiently at high rates. 
In the quantum layer, high-speed and efficient quantum state measurement is generally considered the primary bottleneck~\cite{2016_diamanti_PracticalChallengesQuantum,2017_sasaki_QuantumNetworksWhere,2023_grunenfelder_FastSinglephotonDetectors}. 
For classical data post-processing, the ability to perform high-throughput information reconciliation and privacy amplification is crucial for real-time secret key generation~\cite{2016_diamanti_PracticalChallengesQuantum,2018_yuan_10MbQuantumKey,2023_yang_InformationReconciliationContinuousvariables}. 

\begin{figure*}[!hbt]
    \centering
    \includegraphics[width=0.9\textwidth]{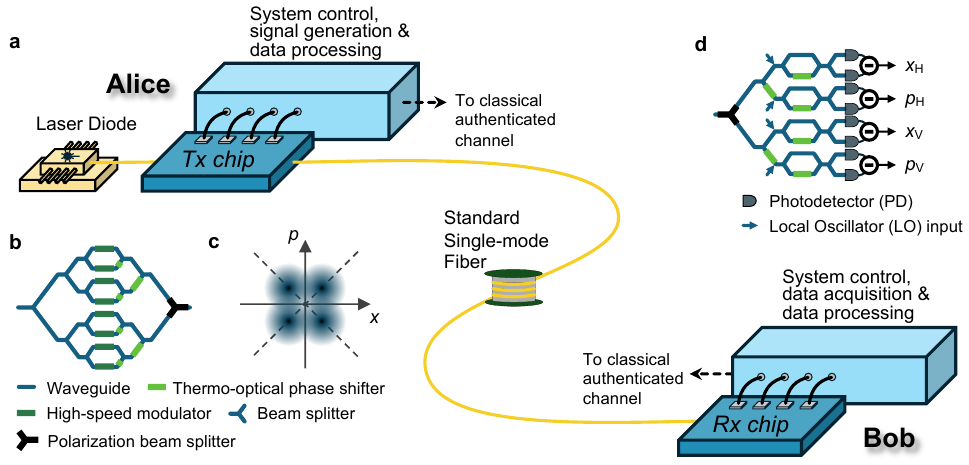}
    \caption{(a) Schematic illustration of our DM CV QKD experiment. (b) Dual-polarization in-phase quadrature (DP-IQ) modulators for quantum state preparation. (c) Constellation diagram of Quadrature Phase Shift Keying (QPSK) modulation. (d) Conjugate homodyne detection for both polarization components. }
    \label{fig:experiment_schematic}
\end{figure*}

On the other hand, enhancing the technology's viability depends on addressing practical aspects -- cost-effectiveness, compactness, and robustness -- thereby introducing additional layers of complexity and making the task more challenging~\cite{2016_diamanti_PracticalChallengesQuantum}. 

In this paper, we address these challenges and report a high-performance, cost-effective quantum key distribution solution based on integrated photonics, achieving a secret key rate of 1.213 $\SI{}{\giga\bit\per\second}$ over a metropolitan distance of $\SI{10}{\km}$.

In the quantum layer, we achieved ultra-high rate quantum state preparation and measurement by leveraging advanced silicon photonics technology and optimizing a wide-bandwidth detector design~\cite{2024_ng_ChipintegratedHomodyneDetection}.  

The adoption of integrated silicon photonics presents several key advantages for our quantum system. 
First, photonic integrated circuits (PICs) achieve strong optical mode confinement, enabling high-density integration of various compact optical components while enhancing the functionality and robustness of the entire quantum optical system.

Second, silicon photonics provides high-speed electro-optical conversion, permitting ultrafast optical signal modulation and detection. Specifically, for quantum state measurement with single-photon energy level input, detector design must meet stringent requirements for both high gain and low noise. These specifications traditionally limit the operation speed of detectors for quantum information processing (QIP) applications. 
To address this challenge, we adopted our modeling approach to develop an optimized detector design with enhanced heterodyne performance, achieving an ultrafast quantum receiver operating at $\SI{40}{Gbaud}/s$ on a single photonic chip at room temperature.

Third, silicon photonics capitalizes on its compatibility with mature CMOS manufacturing processes. This makes it an ideal platform for cost-effective mass production and monolithic integration of optical and electronic components.

With our on-chip quantum transmitter and receiver, we report a chip-based experimental demonstration of a polarization multiplexing continuous-variable quantum key distribution (CV QKD) protocol using discretely modulated coherent states, with composable finite-size security against collective attacks \cite{2023_kanitschar_FiniteSizeSecurityDiscreteModulated}.

In CV QKD protocols, the use of homodyne detectors offers significant advantages for practical applications in terms of cost-effectiveness, integrability, and room-temperature operation~\cite{2009_lvovsky_ContinuousvariableOpticalQuantumstate}. 
Moreover, CV QKD is highly compatible with existing telecommunication technologies, such as high-performance modulators, signal generators, and digital signal processing (DSP)~\cite{2002_grosshans_ContinuousVariableQuantum,2003_grosshans_QuantumKeyDistribution,2018_laudenbach_ContinuousVariableQuantumKey} and - as recently demonstrated - integrates well with the inherent topology of optical networks \cite{bian2023highratepointtomultipointquantumkey, Hajomer_2024, 2024_securitymultiuserDMCV}. Another major advantage of CV QKD is its resistance to Raman scattering noises in optical fibres, making it a promising candidate for integration into existing telecom networks that co-propagate with classical data traffic~\cite{2010_qi_FeasibilityQuantumKey}. 

Discrete-modulated (DM) CV QKD protocols further enhance the technology's practicality by simplifying implementation; they allow quantum state preparation using standard (discrete) modulation formats like QPSK or QAM instead of assuming a (continuous) Gaussian distribution. 
Consequently, this approach reduces the complexity of modulation instruments, random number generation, and information reconciliation
~\cite{2009_leverrier_UnconditionalSecurityProof,2023_leverrier_InformationReconciliationDiscretelymodulated,2024_bauml_SecurityDiscretemodulatedContinuousvariable, jaksch2024composablefreespacecontinuousvariablequantum}. 

Substantial progress has been made in DM CV QKD protocols in recent years, particularly in security analysis, aimed at establishing rigorous and tight lower bounds on secret key rates secure against the most general types of attacks~\cite{2019_lin_AsymptoticSecurityAnalysis,2019_ghorai_AsymptoticSecurityContinuousVariable,2020_lin_TrustedDetectorNoise,2021_kaur_AsymptoticSecurityDiscretemodulation,2021_matsuura_FinitesizeSecurityContinuousvariable,2023_kanitschar_FiniteSizeSecurityDiscreteModulated,2024_bauml_SecurityDiscretemodulatedContinuousvariable,2024_primaatmaja_DiscretemodulatedContinuousvariableQuantum, 2025_Garcia_Improved_rates_for_DMCVQKD}. 
However, experimental investigation of these protocols' practical advantages remains in its early stages. This motivates our research, where we developed chip-based, high-performance quantum transmitter and receiver, and proposed a realization scheme for DM CV QKD with particular focus on the information reconciliation algorithm. 
Our results demonstrate a QKD system that achieves both high key rates and cost efficiency, presenting a practical solution for real-world applications. 

\section{Protocol Description and Secret key rate}
The DM CV QKD protocol and the corresponding security proof, implemented in our experiment and based on Quadrature Phase Shift Keying (QPSK) modulation, are presented below.

\begin{widetext}
\begin{prot} 
\textit{Arguments:}\\
$N$ -- total number of rounds\\
$\beta_\text{ET}$ -- radial threshold for energy test\\
$k_T$ -- number of rounds selected for both the energy test and acceptance test.\\
$M$ -- measurement range of the heterodyne detector\\
$\Delta_r$ -- radial post-selection parameter\\
$\epsilon_\text{cor}$ -- correctness parameter\\
$\epsilon_\text{sec}$ -- secrecy parameter\\
$\epsilon_\text{ET}$, $\epsilon_\text{AT}$, etc.  -- security parameters related to the energy test, the acceptance test, etc.\\[5pt]
\noindent \textit{Protocol:}
\begin{enumerate}
    \item (State preparation) In each round of experiments, Alice prepares a coherent state $\ket{\alpha}$ with $\alpha = \alpha_0 e^{i(x\frac{\pi}{2}+\frac{\pi}{4})}$, where $x \in \{0,1,2,3\}$ is determined by a two-bit random number. Then, Alice sends $\ket{\alpha}$ to Bob via an untrusted quantum channel. 
    \item (State measurement) After receiving the quantum state, Bob performs conjugate homodyne detection, commonly referred to as heterodyne detection in the continuous-variable quantum community, and obtains canonically conjugate quadrature components $p$ and $q$. 
    
    Steps (1) and (2) are repeated $N$ times. 

    \item (Energy test) After running Steps (1) and (2) for a total of $N$ rounds, Bob randomly chooses $k_T$ rounds ($k_T \ll N$) and performs an energy test. 
    The purpose of this energy test is to certify that the quantum state being received by Bob has most of its weight residing within a finite-dimensional Hilbert space, except with some small probability $\epsilon_\text{ET}$. 
    More specifically, this is done by checking if the frequency of the heterodyne measurement results exceeds some predetermined threshold $\sqrt{\frac{p^2 + q^2}{2}} \geq \beta_{\text{ET}}$. If the test succeeds, Alice and Bob proceed; otherwise, they abort. 

    \item (Acceptance test) 
    If the energy test was successful, they disclose the data from the rounds used in the energy test via the classical channel. To ensure composability, before protocol execution, Alice and Bob need to determine an expected channel behavior which is used to define the acceptance set. 
    Based on the disclosed data, Alice and Bob determine statistical estimators for their observables, which are compared to the pre-defined acceptance set. Except with probability $\epsilon_{\mathrm{AT}}$, this ensures the unknown quantum state is considered in the security analysis. In case the observed statistics lie within the set, the test passes, otherwise the protocol aborts.

    \item (Key map) Bob performs a key map on the remaining $n:=N-k_T$ rounds to determine the raw key string $\tilde{z}$. For this purpose, Bob's measurement outcomes are discretized to an element in the set $\{0,1,2,3\}$, where outcomes mapped to $\perp$ are discarded, allowing for postselection. For the QPSK modulation format specifically, the corresponding key map is expressed as
    \begin{equation}
    \tilde{z}_k = \left\{
    \begin{aligned}
        0 &\quad\text{if}\, 0\leq \arg(p_k+q_k) < \frac{\pi}{2} \,\wedge\, \Delta_r\leq \text{abs}(p_k+q_k)\leq M, \\ 
        1 &\quad\text{if}\, \frac{\pi}{2} \leq \arg(p_k+q_k) < \pi \,\wedge\, \Delta_r\leq \text{abs}(p_k+q_k)\leq M, \\ 
        2 &\quad\text{if}\, \pi \leq \arg(p_k+q_k) < \frac{3\pi}{2} \,\wedge\, \Delta_r\leq \text{abs}(p_k+q_k)\leq M, \\ 
        3 &\quad\text{if}\, \frac{3\pi}{2} \leq \arg(p_k+q_k) < 2\pi \,\wedge\, \Delta_r\leq \text{abs}(p_k+q_k)\leq M, \\ 
        \perp &\quad \text{otherwise}. 
    \end{aligned}
    \right.
    \end{equation}
    An illustrative sketch of the key map is shown in Fig.~\ref{fig:key_map}. It should be noted that our QPSK modulation format and key map differ by $\pi/4$ 
    compare to Ref.~\cite{2023_kanitschar_FiniteSizeSecurityDiscreteModulated}. However, this difference can be incorporated in the security analysis easily. 

    \item (Error correction) Alice and Bob publicly communicate over a classical authenticated channel to reconcile their raw keys $\tilde{x}$ and $\tilde{z}$. After the error correction phase, Alice and Bob share a common string except with a small probability $\epsilon_\text{EC}$.
    
    \item (Privacy amplification) Alice and Bob apply a two-universal hash function to their common string. Except with small probability $\epsilon_\text{PA}$, in the end, Alice and Bob hold a secret key. 

\end{enumerate}

\end{prot}

\begin{figure}[ht]
    \centering
    \includegraphics[width=0.35\textwidth]{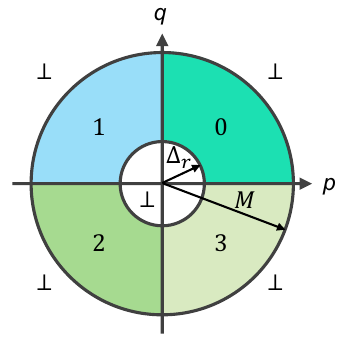}
    \caption{Sketch of the key map in phase space used in experiment. Heterodyne measurement results $p$ and $q$ falling within the region denoted as $\perp$ are discarded, while those falling within the four shaded areas are assigned symbols $y\in\{0,1,2,3\}$.}
    \label{fig:key_map}
\end{figure}
\end{widetext}

\subsection{Security Statement}

We applied the composable security argument from Refs. \cite{Kanitschar_Thesis_2022, 2023_kanitschar_FiniteSizeSecurityDiscreteModulated} which proves security of the present protocol within Renner's composable security framework~\cite{2005_renner_UniversallyComposablePrivacy,2005_renner_SecurityQuantumKey,2014_portmann_CryptographicSecurityQuantum}. Let $\epsilon_{\mathrm{ET}}, \epsilon_{\mathrm{AT}}, \epsilon_{\mathrm{EC}}$ and $\epsilon_{\mathrm{PA}}$ be the security parameters of the subprotocols listed in \textbf{Protocol 1} and let $\bar{\epsilon}>0$ be a virtual parameter describing the applied smoothing. Then, the objective QKD protocol is $\epsilon_\text{cor} = \epsilon_\text{EC}$ correct and $\epsilon_\text{sec}=\max\{\frac{1}{2}\epsilon_\text{PA}+\bar{\epsilon},\epsilon_\text{ET}+\epsilon_\text{AT}\}$ secret, so $\epsilon:=\epsilon_\text{sec}+\epsilon_\text{cor}$ secure against collective independently and identically distributed (i.i.d.) attacks. 
The $\epsilon_\text{cor}$-correctness condition, $\Pr[s_A\neq s_B]\leq\epsilon_\text{cor}$, describes the situation where the protocol does not abort and Alice and Bob do not share the same key. The $\epsilon_\text{sec}$-secrecy condition captures the situation where the protocol does not abort and the shared key is not private, i.e., known to Eve. 

In case the protocol does not abort, the secure key length satisfies~\cite{2023_kanitschar_FiniteSizeSecurityDiscreteModulated}
\begin{equation}
\begin{aligned}
    l\leq & n \Big[\min_{\rho\in\mathcal{S}^\text{E\&A}} H(X|E')_\rho - \delta(\bar{\epsilon}) - \Delta(w)\Big] - \text{leak}_\text{EC}\\
    & - 2\log_2\Big(\frac{1}{\epsilon_\text{PA}}\Big),
\end{aligned}\label{eq:key rate}
\end{equation}
where the minimization of the conditional von Neumann entropy is taken over all states in a finite-dimensional Hilbert space that pass both the energy test and the acceptance test. 
$\delta(\bar{\epsilon})$ and $\Delta(w)$ are penalty terms related to dimension reduction. 
$\text{leak}_\text{EC}$ corresponds to the total information leakage in the error correction (information reconciliation) process, and the term $2\log_2\Big(\frac{1}{\epsilon_\text{PA}}\Big)$ is related to the privacy amplification process. 
We refer readers to the composable finite-size security
proof~\cite{2023_kanitschar_FiniteSizeSecurityDiscreteModulated} for detailed security definitions and analysis.

\section{Experimental Setup \& Photonic Chip Fabrication}
\begin{figure*}[ht]
    \centering
    \includegraphics[width=0.9\textwidth]{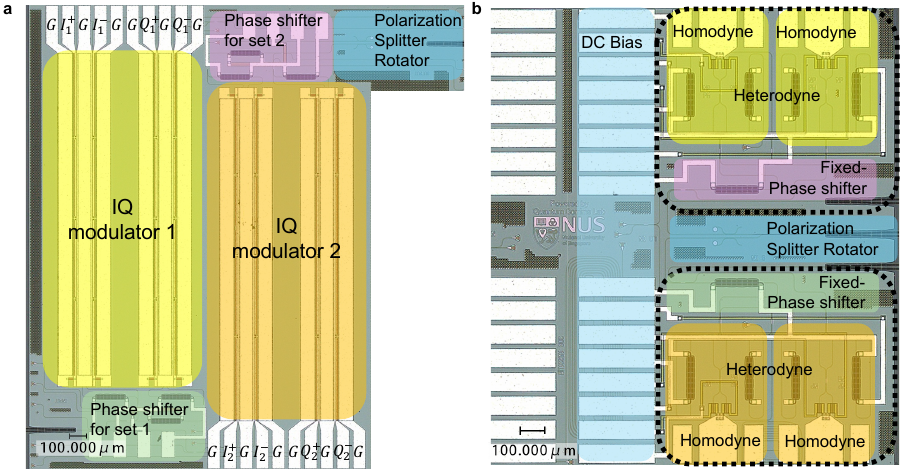}
    \caption{Microscope image of our DM CV QKD chip. (a) Transmitter chip: Two sets of IQ modulators are used for quantum state preparation in the two orthogonal polarization modes, respectively. Thermal-optical phase shifters (TOPS) are used for fine-tuning the relative phase and locking the IQ modulators to their proper working point. A polarization converter then rotates one TE mode to TM mode and combine with the other to work as polarization beam splitter. Finally, the polarization multiplexed signal is coupled to single-mode fiber via a polarization-insensitive edge coupler. (b) Receiver chip: The polarization-multiplexed signals are first converted into two TE modes using an edge coupler and a polarization beam splitter. Two sets of heterodyne detection setups are deployed for conjugate homodyne detection for each polarization mode. Mach-Zehnder Interferometers (MZI) with relative phase tuning are introduced as tunable beam splitters for balanced homodyne detection. Similar to the transmitter chip design, TOPSs are deployed for phase fine-tuning. The subtracted photocurrents are then wired to a PCB and amplified using RF low-noise amplifiers for later data processing. }
    \label{fig:Chip_microscope_image}
\end{figure*}

In our experiment, as depicted in Fig.~\ref{fig:experiment_schematic}, Alice holds the transmitter chip producing dual-polarization QPSK (DP-QPSK) signals $\ket{\alpha} = \ket{\alpha_0 e^{i(x\frac{\pi}{2}+\frac{\pi}{4})}}$ with $x \in \{0,1,2,3\}$. 
Bob holds the receiver chip, which contains two heterodyne detectors for conjugate homodyne detection of both polarization components. 
Between them lies a spool of standard single-mode optical fiber (Corning SMF-28 Ultra), simulating real-world quantum channel for state propagation. 
Both the transmitter and receiver chips are fabricated by Advanced Micro Foundry Singapore using their active silicon-on-insulator (SOI) platform. 
Microscopic photographs of our transmitter and receiver chips are shown in Fig.~\ref{fig:Chip_microscope_image} (a) and (b), respectively.

\subsection{Transmitter} 
In the transmitter, an external cavity laser (ECL) with a linewidth of 50 kHz is coupled into the transmitter chip for quantum state preparation. Within the chip, a 50:50 beam splitter (BS), implemented using a Multi-Mode Interferometer (MMI), first divides the light into two paths. Each path generates a QPSK signal for one polarization, which are then combined at the polarization beam splitter to produce a dual-polarization (DP) QPSK signal (Fig.~\ref{fig:Chip_microscope_image} (a)).

To achieve high-rate and high-fidelity quantum state preparation, we utilize traveling-wave carrier-depletion silicon modulators operating under reverse bias condition (-2 V) for efficient and high-speed signal modulation. 
Additionally, a custom-designed dual-drive push-pull configuration is implemented as Mach-Zehnder modulator (MZM), enabling precise modulation of both the I and Q components. 
In this configuration, the silicon modulators are driven with opposite voltages ($I^+ = -I^-$ and $Q^+ = -Q^-$) with a small peak-to-peak modulation voltage of 0.5 V. This allows us to perform signal modulations in an (almost) linear region, which also reduces the voltage requirement compared to the approach by modulating the half-wave voltage $V_\pi$, and minimizes signal distortions caused by phase-intensity correlation due to plasma dispersion effect in silicon modulators. 

The IQ modulation signals are determined by random numbers generated from our chip-based quantum random number generator (QRNG)~\cite{2024_ng_ChipintegratedHomodyneDetection}. 
After necessary preprocessing (see Section~\ref{sec:DSP} for details), the radio frequency (RF) modulation signals are generated using an Arbitrary Waveform Generator (AWG) (Keysight M8195A) and fed into the transmitter chip via RF probes.

For precise and stable state generation over long periods, our IQ modulators are locked to their desired operational point using on-chip thermo-optic phase shifters (TOPS). Specifically, as illustrated in Fig.~\ref{fig:experiment_schematic} (c), a TOPS is integrated into each MZM, ensuring that it operates at the NULL point. Between the I and Q modulators, an additional TOPS introduces a fixed $\pi/2$ phase difference. Next, an on-chip polarizing beam splitter (PBS) converts one optical mode from TE to TM and combines it with the other optical mode. The combined signal is then coupled to a single-mode optical fiber via an polarization insensitive edge coupler. In addition, variable optical attenuators (VOAs, not shown in the schematic) further attenuate the optical signal to the single-photon energy level, as required by parameter optimization to achieve optimal secret key rate. 

Further details on transmitter performance characterization are provided in Section~\ref{subsect:Transmitter Chip Characterisation}.

\subsection{Receiver}

On the receiver side, the optical signal enters the receiver chip (Fig.~\ref{fig:Chip_microscope_image} (b)) through an edge coupler and is then directed through a polarizing beam splitter to separate the incoming signal into two orthogonal polarization states. 

The signal in each polarization mode is then separated using a BS implemented with an MMI. In one arm, a fixed phase difference of $\pi$/2 is introduced by a TOPS. The signals are then directed to two balanced homodyne detectors, forming a conjugate homodyne detection setup. 
To achieve a precise balancing of homodyne detection on chip, we employ a tunable 50:50 BS scheme consisting of two MMIs and two TOPSs, fabricated to form a Mach-Zehnder Interferometer (MZI). The input signal and the strong local oscillator (LO) are mixed using this MZI-based tunable BS, and the outputs are fed into two identical high-efficiency germanium lateral PIN photodiodes (PDs) with a responsivity of 0.88 A/W. 

The PDs are designed such that their anode and cathode share a common electrode, which outputs the subtracted photocurrent ($i_1 - i_2$). The remaining electrodes provide the necessary reverse bias for the PDs. The subtracted photocurrents are wired-bonded to a printed circuit board (PCB) and amplified using RF low-noise amplifiers (LNAs). The amplified signals are then sampled using an oscilloscope (Tektronix DPO 72004C) for further processing and analysis. The receiver chip is designed to minimize parasitic capacitance, and all RF traces on both the chip and PCB are optimized to improve spectrum flatness and bandwidth, thus enhancing signal integrity. 

Our quantum receiver achieves a common-mode rejection ratio (CMRR) exceeding 20 dB over a wide operating bandwidth up to 20 GHz for each homodyne detection setup. These results confirm that our on-chip PDs exhibit consistent performance, maintain well-balanced operation for balanced homodyne detection, and that the wire bonds and PCB signal traces have a minimal impact on signal integrity.

As the security of the DM CV QKD protocol relies on a trusted detector model, we characterize the receiver efficiency ($\eta_D$) as 0.33, which accounts for the optical coupling loss to the receiver chip, the insertion losses of the BS, and waveguide propagation losses. The electronic noise of our homodyne detection system, including contributions from LNA and oscilloscope, is calibrated to be $\nu_{el}$ = 0.043 (in shot-noise unit). 

Further details on receiver performance characterization are provided in Section~\ref{subsect:Receive Chip Characterisation}. 

\subsection{Digital signal processing (DSP)} \label{sec:DSP}
\begin{figure}[t]
    \centering
    \includegraphics[width=0.4\textwidth]{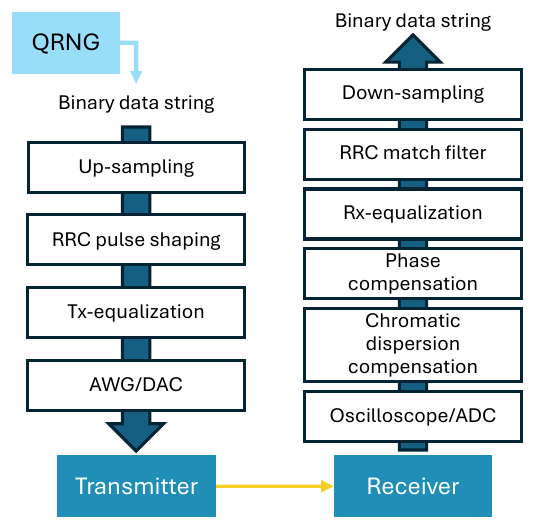}
    \caption{The DSP flowchart of our DM CV QKD experiment.} 
    \label{fig:dsp}
\end{figure} 

Similar to modern digital coherent communication, CV QKD can leverage advanced digital signal processing (DSP) technology to significantly enhance practical performance while reducing hardware complexity~\cite{2010_savory_DigitalCoherentOptical,2023_chen_ContinuousmodeQuantumKey}. 
In this work, we implement DSP algorithms for multiple purposes: Nyquist pulse shaping to maximize spectral efficiency and minimize intersymbol interference (ISI), compensation of channel propagation impairments including chromatic dispersion, phase recovery compensation, and equalization of transmitter and receiver devices. Fig.~\ref{fig:dsp} shows the block diagram of DSP algorithms used in our system. 

After getting the random number sequence, Alice first upsampled the signal system frequency from 40 GSamples/s to 65 GSamples/s to match with the operation rate of the AWG, which functions as a digital-to-analog converter (DAC). A root-raised cosine (RRC) finite impulse response (FIR) filter with a roll-off factor of 0.4 is then applied to create a band-limited baseband signal.  Next, a Tx equalizer is deployed to compensate for transmitter device impairments, including those introduced by AWG, RF probes and silicon modulators. Finally, Alice adds 
pilot pulses with a higher amplitude adjacent to the quantum signal block to facilitate carrier phase recovery. 

On Bob’s side, after quantum state measurement, the output signals are acquired by an oscilloscope, which functions as an analog-to-digital converter (ADC) with a sampling rate of 50 GSamples/s. 
After chromatic dispersion compensation, a phase correction signal is computed based on the measured pilot pulses. 
After phase recovery, the signal is processed through an Rx equalizer to compensate for analog imperfections in the receiver device. Equalizers for both transmitter and receiver are implemented using FIR filters, which are constructed by inverting the frequency response of transmitter and receiver, respectively. 
Subsequently, the equalized signal is processed through a matched RRC filter, which satisfies the Nyquist criterion at the sampling points to minimize ISI. Finally, the signal is downsampled back to 40 GSamples/s to get the final binary data string. 

Only linear DSP algorithms were used in our integrated experimental system, and their parameters were optimized to achieve the best performance.

\section{Classical data post-processing} 

Information reconciliation (IR) is a crucial step in the classical data post-processing phase in QKD, as it corrects any discrepancies between their correlated sifted keys. 
The problem of one-way IR in QKD is equivalent to the problem of source coding with side information in information theory~\cite{2011_elkouss_InformationReconciliationQuantum,2017_tomamichel_FundamentalFiniteKey,2011_elgamal_NetworkInformationTheory}, which can be solved by using error correction (EC) codes, such as low-density parity-check (LDPC) codes, turbo codes and polar codes.

Polar codes present an attractive solution for IR in QKD as it can approach Shannon's capacity limit while providing convenient support for rate adaption~\cite{arikan_channel_2009,babar_polar_2020}. 
Beyond these theoretical advantages, polar codes have demonstrated their practical viability through extensive research and engineering efforts, leading to its adoption in the 5G New Radio (NR) standard and high-throughput (multiple Gb/s) implementations through Application-Specific Integrated Circuit (ASIC) technology~\cite{liu_516gbps_2018,shao_survey_2019}.
Furthermore, polar codes with one-way classical communication bring two benefits to QKD systems: it aligns naturally with established QKD security analysis methods and reduces latency by decreasing communication time, making them suitable for high-rate, long-distance QKD applications. 

In this work, we adopt two approaches to enhance the practical performance of EC based on polar codes in finite block length settings: a list decoding algorithm assisted by cyclic redundancy check (CRC) for block error verification~\cite{tal_list_2015,balatsoukas-stimming_llr-based_2015}, and bit-channel evaluation optimization in the construction of polar code using Monte Carlo simulation~\cite{nakassis_polar_2014}.

In order to evaluate the amount of leaked information $\text{leak}_{EC}$, in compliance with the composable security requirement for QKD security analysis~\cite{2023_kanitschar_FiniteSizeSecurityDiscreteModulated}, we adopt a strategy of leaking all bits in failed EC blocks in IR to the eavesdropper. 
More specifically, after the quantum layer, the energy test, the acceptance test and the key mapping phase, Alice and Bob partition the total number of preserved (sifted) keys, $N_{ECtot}$, into $N_{ECbk}$ EC blocks of length $N_{EC}$ and perform on each EC block separately. 
In the event of unsuccessful EC decoding -- where the processed key strings differ -- Bob reveals the entire EC block to Eve by transmitting it to Alice for rectification through the authenticated classical communication channel. 

The amount of leaked information, $\text{leak}_{EC}$, can therefore be expressed as 
\begin{widetext}
\begin{equation}
\begin{aligned}
    \text{leak}_{EC} =& N_{ECbk}(1-\text{FER})(N_{EC}-K+n_{hash}) + N_{ECbk}\cdot\text{FER}\cdot N_{EC},\\
    =& N_{ECtot} \bigl\{ (1-\text{FER}) \big(1-\beta_{QKD}\cdot C+\frac{n_{hash}}{N_{EC}}\big) + \text{FER} \bigl\}.
\end{aligned}
\end{equation}
\end{widetext}

Here, $C$ corresponds to the Shannon capacity of the binary symmetric channel (BSC) with crossover probability $p$ (which corresponds to quantum bit error rate, QBER), that is, $C = 1- H_{2}(p)$, where $H_{2}(p) = -p\log(p)-(1-p)\log(1-p)$ is the binary Shannon entropy. $\beta_{QKD} = K_{QKD}/(N_{EC}\cdot C)$ denotes the reconciliation efficiency, with $K_{QKD} = K - \text{CRClen}$, where $\text{CRClen}$ represents the length of the CRC parity check bits. 
$\text{FER}$ refers to the frame error rate, which indicates the average failure probability of each EC block. 
$n_{hash}$ denotes the number of bits used for error verification for each processed EC block. 

In our experiment, we configure the IR algorithm parameters with $\text{CRClen} = 8$ and a list size $L = 32$ for the list decoder. 
For a BSC with crossover probability $\text{QBER} = 0.35$, we target reconciliation efficiencies $\beta_{QKD}$ of $\{0.80, 0.85, 0.90, 0.95\}$ and perform numerical simulation 1000 times across various EC block length settings $N_{EC} =\{2^{10}, 2^{12}, 2^{14}, 2^{15}, 2^{16}\}$.
With the assistance of the $\text{FER}$ obtained through simulation, we select the optimal IR parameter setting to minimize information leakage $\text{leak}_{EC}$ for the experiment. 

Following the EC process, we perform privacy amplification using Toeplitz hashing -- a class of functions in the 2-universal hash function family -- to obtain the final secure keys. 
The implementation of Toeplitz hashing has been extensively studied, with efficient schemes leveraging hardware parallelization on FPGAs and GPUs, as well as software-based algorithmic accelerations such as Fast Fourier Transform (FFT). 
These advancements have enabled state-of-the-art implementations to achieve multi-gigabit throughput~\cite{2016_zhang_FPGAImplementationToeplitz,2019_zheng_6GbpsRealtimea,2020_drahi_CertifiedQuantumRandom}.

\section{Results}

\begin{table}[t]
    \centering
    \renewcommand{\arraystretch}{1.2}
    \begin{tabular}{|l|c|c|}
        \hline
        \textbf{Parameter} & \textbf{Symbol} & \textbf{Value} \\
        \hline
        Detection efficiency & $\eta_{D}$ & 0.33 \\
        Electronic noise & $\nu_{el}$ & 0.043 \\
        Channel transmittance coefficient & $\eta_{Ch}$ & 0.63387 \\
        Coherent state amplitude & $|\alpha|$ & 0.85 \\
        Cutoff number & $n_c$ & 15 \\
        Detection limit & $M$ & 7 \\
        ET-parameter & $\beta_{\mathrm{ET}}$ & 5 \\
        Testing ratio & $r_T$ & 0.3 \\
        Fraction of outliers & $l_T/k_T$ & $10^{-8}$ \\
        Weight & $\omega$ & $3.72 \times 10^{-8}$ \\
        Post-selection Parameter & $\Delta_r$ & 0.1 \\
        t-factor & $t_F$ & 1.5 \\
        Sec. parameter Energy Test & $\epsilon_{\mathrm{ET}}$ & $\frac{1}{10} \times 10^{-10}$ \\
        Sec. parameter Acceptance Test& $\epsilon_{\mathrm{AT}}$ & $\frac{7}{10} \times 10^{-10}$ \\
        Smoothing parameter & $\bar{\epsilon}$ & $\frac{7}{10} \times 10^{-10}$ \\
        Sec. parameter Error Correction  & $\epsilon_{\mathrm{EC}}$ & $\frac{1}{5} \times 10^{-10}$ \\
        Sec. parameter Privacy Amplification & $\epsilon_{\mathrm{PA}}$ & $\frac{1}{5} \times 10^{-10}$ \\
        \hline
    \end{tabular}
    \caption{Experimental parameters \& protocol parameters}
    \label{tab:parameters}
\end{table}

To determine the optimal coherent state amplitude ($|\alpha|$) and the achievable secret key rate, we first deployed a $\SI{10}{km}$ single-mode spooled fiber, with a characterized propagation loss of $\SI{1.98}{dB}$. We then measured the system detector parameters, including the detection efficiency ($\eta_D$) and electronic noise ($\nu_{el}$), as well as the transmittance coefficient ($\eta_{Ch}$) in a non-adversarial scenario. The total number of experimental runs was set to $1 \times 10^{9}$ and $1 \times 10^{10}$, respectively, with total security parameter $\epsilon=1 \times 10^{-10}$, testing ratio $r_T = 0.3$, and reconciliation efficiency $\beta = \{0.80, 0.85, 0.90, 0.95\}$ under the assumption of zero frame error rate $(\text{FER} = 0)$. This input was used to numerically simulate the achievable secure key rates under the Gaussian channel assumption, allowing us to estimate the optimal coherent state amplitude $|\alpha|$ for the actual protocol execution.

As outlined in the protocol description, to ensure the composability of the statistical testing procedure, it is necessary to perform an acceptance test. Therefore, before the protocol is executed, Alice and Bob need to agree on an expected channel behavior. This could be based on an elaborate channel model, depending on measured system parameters, or by carefully characterizing the system in a non-adversarial scenario. In this work, we chose to characterize the channel under controlled conditions. Based on this characterization, the optimized value of $|\alpha|$ was determined to be 0.85, as derived in the previous step. We then prepared $N= 1 \times 10^{10}$  signals, which were transmitted through the quantum channel, measured by Bob using heterodyne detection and processed through the corresponding DSP algorithms. The obtained values for $q$ and $p$ were used to determine $\langle \hat{n}\rangle_{\beta_i}$ and $\langle \hat{n}^2\rangle_{\beta_i}$, which represent the expectations of the displaced photon number operator and the displaced squared photon number operator, respectively. The experimental results of $\langle \hat{n}\rangle_{\beta_i}$ and $\langle \hat{n}^2\rangle_{\beta_i}$ are presented in Section~\ref{subsect:non-unique acceptance scenario}, which then were used to define the acceptance set $\mathcal{O}$ as the set of all measurement outcomes that have distance of less than $t_X$ where $X \in\{\hat{n}, \hat{n}^2\}$ from the determined values (see Refs. \cite{Upadhyaya_2021,2023_kanitschar_FiniteSizeSecurityDiscreteModulated} for details regarding the definition and Refs. \cite{Upadhyaya_Thesis_2021,hajomer2024experimentalcomposablekeydistribution} for calculation from experimental data).

As for different observables $X$, $t_X$ can be chosen differently, we measured $t_X = t_F \mu_X$ in multiples $t_F$ of the corresponding $\mu_X$ (see Theorem 4 in Ref. \cite{2023_kanitschar_FiniteSizeSecurityDiscreteModulated}) 
\begin{equation*}
    \mu_X := \sqrt{\frac{x^2}{2m_{X}} \ln\left( \frac{2}{\epsilon_{\mathrm{AT}}} \right)}.
\end{equation*}
The case of $t_F = 0$ corresponds to the unique acceptance scenario, which is dominant in literature, while $t_F > 0$ is referred to as non-unique acceptance, allowing for small fluctuations in the measurement result. 

\begin{figure}[t]
    \centering
    \includegraphics[width=0.45\textwidth]{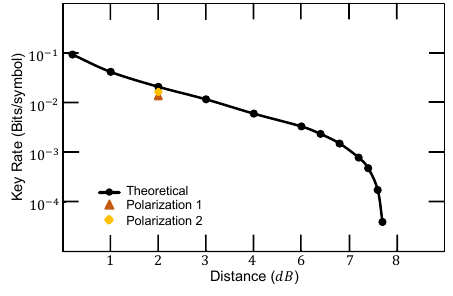}
    \caption{Secure key rate (SKR) versus optical transmission distance. The black curves represent the theoretical SKR assuming a total security parameter of $\epsilon = 1 \times 10^{-10}$, testing ratio $r_T = 0.3$, reconciliation efficiency $\beta = 0.80$, zero frame error rate, and a block size of $1 \times 10^{10}$. The data points for Polarization 1 and Polarization 2 represent the experimentally achieved results.} 
    \label{fig:keyrate}
\end{figure}

Next, through numerical optimization, we determined the optimal radial postselection parameter $\Delta_r$ \cite{lin_asymptotic_2019, Kanitschar_2022} for each polarization setup measurement. Based on the composable finite-size security analysis of our DM CV QKD system, and conditioned on the successful completion of all analytical tests and error correction, we obtained a unique-acceptance secret key rate per symbol of $1.37 \times 10^{-2}$ for $Polarization~1$  and $1.66 \times 10^{-2}$ for $Polarization~2$ with a total security parameter of $\epsilon = 1\times 10^{-10}$. 
This corresponds to a final secret key rate of 1.213 $\SI{}{\giga\bit\per\second}$ using our CV QKD system operating at 40 Gbaud/s with polarization multiplexing. The parameters used in our experiment and in the optimization are listed in Table~\ref{tab:parameters}. While the security parameters for the Energy Test, Acceptance Test, and the smoothing parameter can be chosen directly as given in the table, the given $\epsilon_{\mathrm{EC}}$ and $\epsilon_{\mathrm{PA}}$ serve as upper bounds on the actual parameters, which are determined by the chosen hash-lengths of the respective subroutines.
Our experimental results, along with the simulation results, are presented in Fig.~\ref{fig:keyrate}. 
In addition to the unique-acceptance rate, which is directly comparable to values that can be commonly found in literature, we additionally provide key rates for the non-unique acceptance scenario for $t_F = 1.5$ in Appendix~\ref{subsect:non-unique acceptance scenario}.

\section{Discussion}
In this work, we provide a high-performance and cost-effective QKD system based on integrated silicon photonics, and use it to faithfully run a DM CV QKD protocol with composable finite-size security against collective attacks, achieving a secret key rate of 1.213 $\SI{}{\giga\bit\per\second}$ over a 10 km fiber link with DP-QPSK modulation. 

In the quantum optics layer, we develop a high-speed, silicon photonics-based transmitter and receiver, where all essential components, excluding the laser source, are fully integrated on silicon photonic chips. 
The key component of this system is the high-speed, integrated balanced homodyne detection system, which is designed based on the methodology presented in our earlier work~\cite{2024_ng_ChipintegratedHomodyneDetection}. 
Our integrated photonic platform, combined with optimized DSP techniques, supports on-chip quantum state preparation and measurement at ultra-high rate of 40 Gbaud/s, significantly advancing the practical performance of quantum communication systems. 

In the classical data processing layer, we implement a polar code-based error correction code for efficient information reconciliation. More specifically, we adopt CRC-aided SCL decoder together with bit-channel evaluation optimization for code construction to enhance the efficiency of polar codes in finite block-length settings, thereby limiting the information leakage to adversaries. More importantly, multiple Gb/s implementations of polar code have been reported based on ASIC technology, providing a promising approach for a fully integrated QKD system with high-throughput IR for real-time classical data post-processing.

In Table~\ref{tab:qkd_comparison}, we present a comparison of our work with existing high-rate QKD works. 
Our work demonstrates the first experimental QKD system to achieve the highest symbol rate of 40G, as well as a record-high secret key rate using silicon photonic chips operating at room temperature. 
Our proposed scheme delivers both high performance and cost-effectiveness: it provides a practical QKD solution for real-world problems requiring long-term security, while simultaneously opening new opportunities for applications in cybersecurity networks and Internet-of-Things. 

We also identify several potential directions that could further enhance the performance and security of our system. 
First, adopting higher-order modulation formats such as 16-QAM or 256-QAM could increase the secret key rate per symbol. However, the numerical method we used in Ref.~\cite{2023_kanitschar_FiniteSizeSecurityDiscreteModulated} becomes more computationally challenging as the dimension of the optimization problem increases significantly. Additionally, due to the smaller sample size per symbol, we would expect worse finite-size statistics. 
Second, optimizing on-chip device design could further enhance overall receiver efficiency. From the state-of-the-art results, the minimum insertion loss can be as low as 1.66 dB, including edge coupler, PBSR and 90 degree hybrid, together with a high efficiency PD with efficiency above 92.4\% at 1550 nm, the total efficiency can increase to more than 63.0\%~\cite{pu2010ultra,guan2017compact,dai2018advanced,benedikovic2019comprehensive}. 
This will improve the key rate that equivalent to a 15 km transmission distance reduction, assuming a propagation loss of 0.18 dB/km.
Third, higher-efficiency EC schemes would further reduce information leakage during IR, which is particularly crucial for long-distance DM CV QKD systems. 
Furthermore, our system can readily incorporate advances in security proof techniques, which could provide a higher level of security by defending against coherent attacks~\cite{2024_bauml_SecurityDiscretemodulatedContinuousvariable,2024_primaatmaja_DiscretemodulatedContinuousvariableQuantuma,2025_pascual_garcia_ImprovedFinitesizeKey}, although we note that Refs.~\cite{2024_bauml_SecurityDiscretemodulatedContinuousvariable,2025_pascual_garcia_ImprovedFinitesizeKey} require an additional assumption about the maximum photon number.

\begin{table*}[htbp]
    \centering
    \renewcommand{\arraystretch}{1.2} 
    \setlength{\tabcolsep}{4pt} 
    \small
    \begin{tabularx}{1\textwidth} { 
      | >{\centering\arraybackslash}X 
      | >{\centering\arraybackslash}X 
      | >{\centering\arraybackslash}X 
      | >{\centering\arraybackslash}X 
      | >{\centering\arraybackslash}X 
      | >{\centering\arraybackslash}X 
      | >{\centering\arraybackslash}X 
      | >{\centering\arraybackslash}X 
      | >{\centering\arraybackslash}X | }
        \hline
        \textbf{Ref} & \textbf{QKD Protocol} & \textbf{Security} & \textbf{Attack Level} & \textbf{Post-Processing (IR)} & \textbf{Distance (km)} &  \textbf{Symbol Rate} & \textbf{Secret Key Rate} \\
        \hline
        This work & DM CV (DP-QPSK) & Composable finite-size & Collective & Polar code & 10 & 40G & 1.213 Gbps\\
        \hline
        \cite{hajomer2024experimentalcomposablekeydistribution}(2024) & DM CV (QPSK) & Composable finite-size & Collective & LDPC & 10 & 125M & 1.38 Mbps \\
        \hline
        \cite{hajomer2024continuous}(2024) & DM CV (16-QAM) & Finite-size\footnotemark[1] & Collective & - & 10 & 10G & 351 Mbps \\
        \hline
        \cite{2024_roumestan_ShapedConstellationContinuous}(2024) & DM CV (DP-256-QAM) & Finite-size\footnotemark[1] & Collective & - & 9.5 & 600M & 91.8 Mbps\\
        \hline
        \cite{2024_wang_HighKeyRate}(2024) & GM CV & Asymptotic & Collective & - & 20 & 500M & 10.37 Mbps\\
        \hline
        \cite{2023_tian_HighperformanceLongdistanceDiscretemodulation}(2023) & DM CV (16-APSK) & Asymptotic & Collective & - & 25 & 2.5G & 49.02 Mbps\\
        \hline
        \cite{li2023high}(2023) & Decoy BB84 & Composable finite-size & Coherent & Cascade & 10 & 2.5G & 115.8 Mbps \\
        \hline
    \end{tabularx}
    \footnotetext[1]{The finite-size analysis is only based on worst-case channel parameter estimations.}
    \caption{Performance of our DM CV QKD system as compared to existing high-rate QKD works. QPSK: Quadrature Phase-Shift Keying. QAM: Quadrature Phase-Shift Keying. APSK: Amplitude- and Phase-Shift Keying. DP: Dual-polarization. IR: information reconciliation. LDPC: Low-Density Parity-Check codes. } 
    \label{tab:qkd_comparison}
\end{table*}

\section{acknowledgment}
This research is supported by National University of Singapore, under the Start-up Grant (FY2023). 
F.K. acknowledges funding from the European Union's QuantERA II Programme, Grant Agreement No 101017733, and from the Austrian Research Promotion Agency (FFG), project number FO999891361.

\clearpage
\newpage
\section{Appendix:}
\label{sect:appendix}

\subsection{Transmitter Chip Characterization}

\label{subsect:Transmitter Chip Characterisation}

\begin{figure}[htbp]
    \centering
    \includegraphics[width=0.48\textwidth]{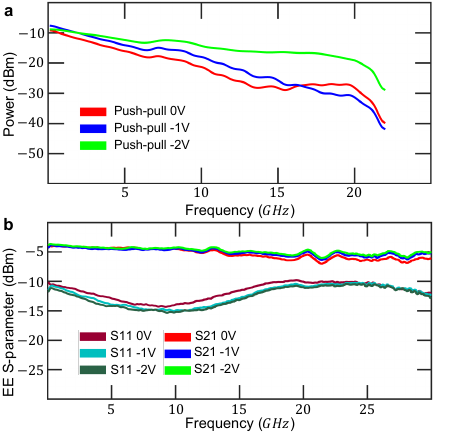}
    \caption{(a) Power spectrum of the transmitter under the push-pull configuration and single-arm operation with different bias voltages. (b) Electrical-to-electrical (EE) S-parameters of the modulator at different bias voltages.} 
    \label{fig:transmitter_characterisation}
\end{figure}

To evaluate the high-speed performance of our on-chip modulator, we performed a series of experimental characterizations using an external photodetector (Conquer KG-PD, 20 GHz) and a spectrum analyzer (Rohde \& Schwarz FSV40). The modulator was tested under a push-pull configuration with varying reverse bias voltages, as shown in Fig.~\ref{fig:transmitter_characterisation}(a). 
The results indicate that at 20 GHz, the transmission loss was reduced from -18 dB (no bias) to -11 dB under a reverse bias of 2 V.

In Fig.~\ref{fig:transmitter_characterisation}(b), we present the results of S-parameter on two ends of the traveling wave electrodes measurements obtained under a 2 V reverse bias using a vector network analyzer (Rohde \& Schwarz ZVA). The measured return loss ($S_{11}$) remained below -10 dB across the frequency range, indicating low reflection and confirming that the modulator is impedance-matched up to 20 GHz. The transmission loss ($S_{21}$) was measured at -5.1 dB at 20 GHz, suggesting low microwave attenuation and confirming efficient high-speed performance.

\subsection{Receiver Chip Characterization}
\label{subsect:Receive Chip Characterisation}

\begin{figure}[htbp]
    \centering
    \includegraphics[width=0.48\textwidth]{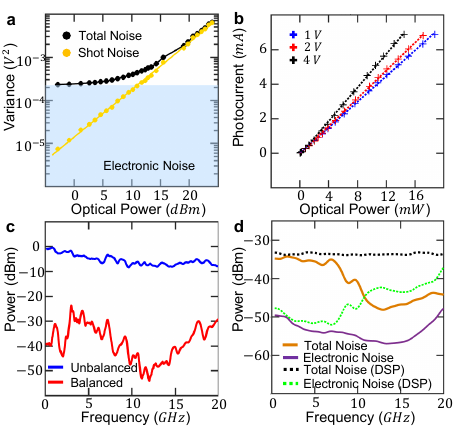}
    \caption{(a) Shot noise as a function of LO power, showing a maximum clearance of $>$13 dB. (b) Photocurrent as a function of input optical power at different reverse bias voltages. (c) Detector balance test demonstrating a common mode rejection ratio (CMRR) of $>$20 dB up to 20 GHz. (d) Comparison of shot noise and electronic noise spectrum before and after DSP application.} 
    \label{fig:receiver_characterisation}
\end{figure}

The characterization results of our quantum receiver are presented in Fig.~\ref{fig:receiver_characterisation}. Figure~\ref{fig:receiver_characterisation} (a) shows the noise characterization of our homodyne detection setup as a function of increasing LO power, demonstrating shot-noise-limited homodyne detection at high LO powers. The shot-noise variance maintains a linear relationship with the LO power, ensuring that the measurement remains accurate and devoid of any nonlinear attributes. 
The quantum receiver is measured to saturate at approximately 23.6 dBm LO power (before coupling), with a quantum-to-classical noise ratio (QCNR) of 14.3 dB. 
Figure~\ref{fig:receiver_characterisation} (b) illustrates the photodiode responsivity under various reverse bias conditions. At 1 V and 2 V reverse bias, our on-chip PDs exhibit high linearity with increasing LO power, demonstrating stable performance across the tested conditions.
Figure~\ref{fig:receiver_characterisation} (c) represents the noise characterization in both balanced and unbalanced configurations. The differences between these measurements indicate a common-mode rejection ratio (CMRR) of our balanced homodyne detectors exceeding 20 dB over a wide operating bandwidth up to 20 GHz. These results confirm that our on-chip PDs exhibit consistent performance, maintain well-balanced operation for balanced homodyne detection, and that the wire bonds and signal traces on PCB minimally impact signal integrity. 
The measured shot-noise spectrum is shown in Fig.~\ref{fig:receiver_characterisation} (d). Although the 3-dB bandwidth is approximately 7.8 GHz, the shot-noise-limited measurement extends beyond 22 GHz (not shown in the figure). The figure also demonstrates that with DSP, the frequency response of the detector can be improved digitally, with the shot-noise clearance remaining unchanged. Table~\ref{tab:optical_parameters} presents the optical loss of each individual on-chip optical component used in the experiment.

\begin{table}[t]
    \centering
    \renewcommand{\arraystretch}{1.2}
    \resizebox{0.45\textwidth}{!}{ 
        \begin{tabular}{|l|c|}
            \hline
            \textbf{Optical Component} & \textbf{Insertion Loss (dB)} \\
            \hline
            Grating Coupler & 4.54 \\
            Edge Coupler & 2.88 \\
            Thermo-optical Tunable Phase Shifter & $<$0.2 \\
            Multi-Mode Interferometer (MMI) & $<$0.2 \\
            MZ Modulator & 3.3 \\
            Polarization Beam Rotator-Splitter (PBRS) & 0.8 \\
            Waveguide (1 cm) & 2.0 \\
            \hline
        \end{tabular}
    }
    \caption{Insertion loss of optical components.}
    \label{tab:optical_parameters}
\end{table}

\subsection{Details of polar code-based information reconciliation}
\label{subsect:polarcode}

\begin{figure*}[t]
    \centering
    \includegraphics[width=0.65\textwidth]{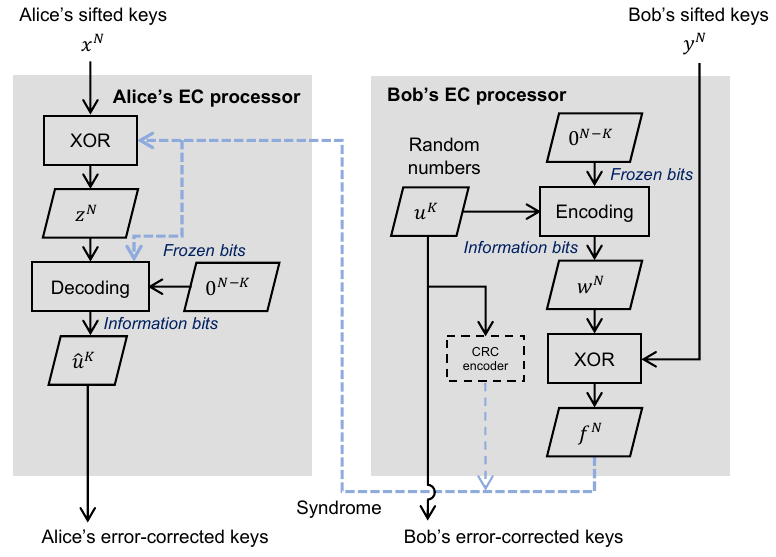}
    \caption{The workflow of the IR scheme based on polar codes. 
    We adopt a reverse reconciliation scheme where Alice corrects her keys to align them with Bob's keys. 
    CRC encoders are used to generate parity check bits for block error verification. 
    The list decoding architecture and polarization channel optimization are not depicted in this figure. Information leaked by the CRC parity check bits will be eliminated in the subsequent privacy amplification phase. } 
    \label{fig:polarcodeworkflow}
\end{figure*}

In our experimental DM CV QKD system, we adopt the implementation scheme of polar codes from~\cite{nakassis_polar_2014} for information reconciliation. The workflow of the IR scheme is illustrated in Fig.~\ref{fig:polarcodeworkflow}.

In our setup, we use the polarization weight (PW) algorithm~\cite{he_beta-expansion_2017,chen_investigation_2018} in conjunction with Monte Carlo-based channel optimization to construct polar codes. Moreover, we use the CRC-assisted SCL algorithm~\cite{tal_list_2015} to enhance the decoder's performance during the decoding process. In the SCL algorithm, the decoder does not simply decode the message bits sequentially; instead, it concurrently considers $L$ most probable decoding paths at each decoding stage. 
In the final decoding stage, the most likely path is selected, aided by CRC parity check bits. 

Numerical simulations are performed to evaluate the performance of the proposed polar code-based EC scheme using randomly generated sifted keys. 
We tested a wide range of QBER, ranging from 10\% to 35\%. The EC block length $N = 2^n$ is selected for $n$ values from 10 to 16. 
The CRC length $\text{CRClen}$ is chosen from the set $\{8, 16\}$, and the list size for the list decoder ranges from 1 (corresponding to the original SC decoder case) to a maximum of 128. Moreover, we set a target reconciliation efficiency of $\{80\%, 85\%, 90\%, 95\%\}$. 

For performance evaluation, we assess the Frame Error Rate (FER), which represents the overall probability that the two key strings differ after the EC process. The simulation results are provided in Table~\ref{tab:ECfinalresulttable}.

\begin{table*}
\caption{\label{tab:ECfinalresulttable} Simulation results for the EC scheme based on polar code. The CRC length $\text{CRClen}$ is set to 8, and the list size $L$ is set to 32. 
}
\begin{tabular}{|wc{6em}|wc{4em}|wc{4em}|wc{4em}|wc{4em}|wc{4em}|wc{4em}|} 
\hline
QBER = 0.1 & \multicolumn{3}{c|}{$\beta_{\rm QKD}$ = 80\%} & \multicolumn{3}{c|}{$\beta_{\rm QKD}$ = 85\%}\\
\hline
$\log_2 N$ & 10 & 12 & 14 & 10 & 12 & 14 \\
\hline
FER & 6/1000 & 0/1000 & 0/1000  & 71/1000 & 2/1000 & 0/1000 \\
\hline
QBER = 0.1 & \multicolumn{3}{c|}{$\beta_{\rm QKD}$ = 90\%} & \multicolumn{3}{c|}{$\beta_{\rm QKD}$ = 95\%}\\
\hline
$\log_2 N$ & 10 & 12 & 14 & 10 & 12 & 14 \\
\hline
FER & 254/1000 & 205/1000 & 71/1000 & 534/1000 & 754/1000 & 821/1000 \\
\hline
\hline
QBER = 0.2 & \multicolumn{3}{c|}{$\beta_{\rm QKD}$ = 80\%} & \multicolumn{3}{c|}{$\beta_{\rm QKD}$ = 85\%}\\
\hline
$\log_2 N$ & 10 & 12 & 14 & 10 & 12 & 14 \\
\hline
FER & 61/1000 & 24/1000 & 2/1000 & 190/1000 & 144/1000 & 57/1000 \\
\hline
QBER = 0.2 & \multicolumn{3}{c|}{$\beta_{\rm QKD}$ = 90\%} & \multicolumn{3}{c|}{$\beta_{\rm QKD}$ = 95\%}\\
\hline
$\log_2 N$ & 10 & 12 & 14 & 10 & 12 & 14 \\
\hline
FER & 337/1000 & 453/1000 & 481/1000 & 531/1000 & 773/1000 & 922/1000 \\
\hline
\hline
QBER = 0.3 & \multicolumn{3}{c|}{$\beta_{\rm QKD}$ = 80\%} & \multicolumn{3}{c|}{$\beta_{\rm QKD}$ = 85\%}\\
\hline
$\log_2 N$ & 10 & 12 & 14 & 10 & 12 & 14 \\
\hline
FER & 101/1000 & 95/1000 & 44/1000 & 182/1000 & 235/1000 & 254/1000 \\
\hline
QBER = 0.3 & \multicolumn{3}{c|}{$\beta_{\rm QKD}$ = 90\%} & \multicolumn{3}{c|}{$\beta_{\rm QKD}$ = 95\%}\\
\hline
$\log_2 N$ & 10 & 12 & 14 & 10 & 12 & 14 \\
\hline
FER & 303/1000 & 474/1000 & 651/1000 & 395/1000 & 688/1000 & 919/1000 \\
\hline
\end{tabular}

\begin{tabular}{|wc{6em}|wc{4.9em}|wc{4.9em}|wc{4.9em}|wc{4.9em}|wc{4.9em}|} 
\hline
QBER = 0.35 & \multicolumn{5}{c|}{$\beta_{\rm QKD}$ = 80\%} \\
\hline
$\log_2 N$ & 10 & 12 & 14 & 15 & 16 \\
\hline
FER & 107/1000 & 123/1000 & 98/1000 & 62/1000 & 20/1000  \\
\hline
QBER = 0.35 & \multicolumn{5}{c|}{$\beta_{\rm QKD}$ = 85\%} \\
\hline
$\log_2 N$ & 10 & 12 & 14 & 15 & 16 \\
\hline
FER & 131/1000 & 253/1000 & 327/1000 & 288/1000 & 238/1000  \\
\hline
QBER = 0.35 & \multicolumn{5}{c|}{$\beta_{\rm QKD}$ = 90\%} \\
\hline
$\log_2 N$ & 10 & 12 & 14 & 15 & 16 \\
\hline
FER & 206/1000 & 405/1000 & 636/1000 & 698/1000 & 746/1000  \\
\hline
QBER = 0.35 & \multicolumn{5}{c|}{$\beta_{\rm QKD}$ = 95\%} \\
\hline
$\log_2 N$ & 10 & 12 & 14 & 15 & 16 \\
\hline
FER & 323/1000 & 602/1000 & 860/1000 & 946/1000 & 981/1000  \\
\hline
\end{tabular}

\end{table*}

\subsection{Non-unique acceptance scenario}
\label{subsect:non-unique acceptance scenario}
In the main text, to ensure comparability with the literature, we reported key rates conditioned on acceptance of all analytical tests for the so-called unique acceptance scenario ($t_F = 0$). 
While such key rates are widely reported in the literature, they are difficult to attain in long-term, real-world deployments. This is because each protocol run requires the observed statistics to fall within a strict discrete acceptance set, which is impractical in practice. Additionally, recalculating the key rate based on changing channel behavior is also not feasible for continuous operation.
The security proof method from Ref. \cite{2023_kanitschar_FiniteSizeSecurityDiscreteModulated} provides a solution by allowing the extension of the acceptance set, parametrized by $t_F > 0$, contributing to the completeness of the protocol. In principle, before running the protocol, Alice and Bob need to fix an expected channel behavior. Since predicting the channel via a physical model can be a challenging task, a valid alternative is to characterize the channel in the honest behavior and use the characterization as a reference for the expectations. The sole remaining task when executing the protocol is then checking if the observed statistics are close (quantified by $t_F$) to the expected behavior. 
In what follows, we set $t_F = 1.5$ and demonstrate, based on the results of \textit{Polarization} 1 and \textit{Polarization} 2 shown in Table~\ref{tab:coherent_amplitudes}, how the acceptance test in that case could be executed. We used \textit{Polarization} 2 to characterize the channel in the honest behavior, which allowed us to define the relaxed acceptance set. Assuming \textit{Polarization} 1 was an independent run, we compared our observations with the accepted statistics and found all observables lying within the acceptance set, which means the acceptance test passed successfully. In that way, we achieved a secret key rate per symbol of $1.039 \times 10^{-2}$ under the non-unique acceptance setting.

\begin{table}[htbp]
\centering
\renewcommand{\arraystretch}{1.2}
\begin{tabular}{|c|c|c|c|}
\hline
\textbf{State}& \textbf{Polarization 2}  & \textbf{Polarization 1}  & \textbf{Acceptance Decision}\\   & $\langle\hat{n}_{\beta_i}\rangle_{\mathrm{obs}}^{\mathrm{Pol 2}}$ & $\langle\hat{n}_{\beta_i}\rangle_{\mathrm{obs}}^{\mathrm{Pol 1}}$ & $\langle\hat{n}_{\beta_i}\rangle_{\mathrm{obs}}^{\mathrm{Pol 1}} \leq \langle\hat{n}_{\beta_i}\rangle_{\mathrm{obs}}^{\mathrm{Pol 2}} + t_F \mu_{\hat{n}}$\\
\hline
$\ket{\alpha_0}$ & $2.48 \times 10^{-4}$ & $1.95 \times 10^{-3}$ & \checkmark \\
$\ket{\alpha_1}$ & $7.15 \times 10^{-4}$ & $1.26 \times 10^{-3}$& \checkmark  \\
$\ket{\alpha_2}$ & $3.17 \times 10^{-3}$ & $3.08 \times 10^{-3}$& \checkmark  \\
$\ket{\alpha_3}$ & $2.68 \times 10^{-3}$ & $3.18 \times 10^{-3}$& \checkmark  \\
\hline
\end{tabular}

\vspace{1.5em}

\begin{tabular}{|c|c|c|c|}
\hline
\textbf{State}& \textbf{Polarization 2}  & \textbf{Polarization 1}  & \textbf{Acceptance Decision}\\   & $\langle\hat{n}^2_{\beta_i}\rangle_{\mathrm{obs}}^{\mathrm{Pol 2}}$ & $\langle\hat{n}^2_{\beta_i}\rangle_{\mathrm{obs}}^{\mathrm{Pol 1}}$ & $\langle\hat{n}^2_{\beta_i}\rangle_{\mathrm{obs}}^{\mathrm{Pol 1}} \leq \langle\hat{n}^2_{\beta_i}\rangle_{\mathrm{obs}}^{\mathrm{Pol 2}} + t_F \mu_{\hat{n}^2}$\\
\hline
$\ket{\alpha_0}$ & $1.76 \times 10^{-3}$ & $2.85 \times 10^{-3}$ & \checkmark \\
$\ket{\alpha_1}$ & $4.28 \times 10^{-4}$ & $2.71 \times 10^{-4}$& \checkmark  \\
$\ket{\alpha_2}$ & $3.09 \times 10^{-3}$ & $2.95 \times 10^{-3}$& \checkmark  \\
$\ket{\alpha_3}$ & $3.42 \times 10^{-3}$ & $5.78 \times 10^{-3}$& \checkmark  \\
\hline
\end{tabular}

\caption{Acceptance decision for the non-unique acceptance scenario. We use \textit{Polarization 2} to define the expected channel behaviour in the honest implementation and compare the observations from \textit{Polarization 1} with the acceptance set defined by the expected channel behaviour.} 
\label{tab:coherent_amplitudes}
\end{table}

\clearpage

\bibliography{apssamp}

\end{document}